# Missile guidance law design based on free-time convergent error dynamics

LIU Yuanhe [1, 2], XIE Nianhao[1, 2], LI Kebo[1, 2], LIANG Yangang [1, 2]

1. College of Aerospace Science and Engineering, National University of Defense Technology, Changsha 410072, China
2. Hunan Key Laboratory of Intelligent Planning and Simulation for Aerospace Missions, Changsha 410072, China

**Abstract:** The design of guidance law can be considered a kind of finite-time error-tracking problem. A unified free-time convergent guidance law design approach based on the error dynamics and the free-time convergence method is proposed in this paper. Firstly, the desired free-time convergent error dynamics approach is proposed, and its convergent time can be set freely, which is independent of the initial states and the guidance parameters. Then, the illustrative guidance laws considering the leading angle constraint, impact angle constraint, and impact time constraint are derived based on the proposed free-time convergent error dynamics respectively. The connection and distinction between the proposed and the existing guidance laws are analyzed theoretically. Finally, the performance of the proposed guidance laws is verified by simulation comparison.

**Keywords:** guidance design; free-time convergence; error dynamics approach; impact angle constraint; impact time constraint.

## 1. Introduction

Modern warfare puts forward higher performance requirements and more terminal constraints for tactical missile guidance, which results in the emergence of advanced guidance laws, such as impact angle control guidance (IACG) law and impact time control guidance (ITCG) law. IACG was first proposed by Kim et al.[1] in 1973 and ITCG first appears in 2006 to solve the salvo attack of the anti-ship missile[2]. The mainstream method of IACG and ITCG include improved biased proportional navigation guidance (BPNG) laws[3]-[6], optimal guidance (OG) laws based on various optimal theories[7]-[9], and nonlinear guidance laws based on various modern nonlinear control theories[10]-[13].

*Corresponding author.

Essentially, the guidance law design for an aerodynamically controlled missile is a kind of finite-time error-tracking problem. If the convergence time is too short, the acceleration command, generated by the designed guidance law, may exceed the saturation of the missile actuator, leading to the degradation of guidance performance in application. On the contrary, if the convergence time is too long, there is a risk that the tracking error cannot converge to 0 in the desired time. Therefore, the key to missile guidance design is to make the tracking error converge to 0 in a limited time without generating too largeguidance command. That is, it is necessary to find an appropriate finite-time reaching law.

Some finite-time convergence (FnTC) theories are investigated to find the better reaching law referring to the guidance problem. The FnTC theory was first proposed in[14], whose upper bound of convergence time depends on the initial system states, which need different guidance parameters in different scenarios. Theauthors in [15] proposed an FnTC control approach, where the convergence time can be preset regardless of the initial states and control parameters. While its preset time-varying function is too complex. As an improvement of FnTC, the fixed-time convergence (FxTC) theory is developed in [16], whose upper bound of convergence time is independent of the initial system states. However, the actual convergence time cannot be obtained and there are too many control parameters with unclear physical meanings. To further enhance the constraint on the convergence time, the concept of free-time convergence (FeTC) was proposed in [17], which could be considered as a further development of the FnTC and FxTC. The convergence time is not only independent of the initial system states but also could be set directly according to the physical demand, where the physical meaning of the parameters is clear for a known system. As a result, the FeTC approach is used in the guidance design in this paper.

Unlike general control problems, the prominent feature of guidance law design is to score a hit at the terminal time. A well-known guidance law design mentality is the prediction-correction approach. That is, the terminal state error states are obtained by predicting the terminal system states from the current system states under some heuristics guidance strategy or without any guidance strategy, namely prediction. Then the terminal errors could converge to 0 by adopting advanced control theories, namely correction. For the missile guidance design problem, considering the physical significance of the tracking error in practice, a unified guidance law design approach called optimal error dynamics (OED) was proposed in [18]. The convergence time of the OED approach proves finite but unadjustable by introducing the time-to-go estimation and Schwartz inequality. Furthermore, the OED approach is extended to impact-time-and-angle-control guidance (ITACG) law design in [19]-[20]. Considering the finite-time convergence of the guidance tracking error, a finite-time convergent sliding mode guidance law was proposed in [21]-[22], which is an application of the FnTC approach in [14]. Similar to the OED in [18], a unified guidance law design approach called fixed-time convergent error dynamics (FxTCED) was proposed in [23] by using the FxTC approach in [16]. The specific design process of IACG and ITCG is given as examples. The OED and FxTCED approaches could be further extended to multi-constraint cooperative guidance and intercepting maneuvering targets[24]. The finite time control theory (FTC), which is closely related to the FnTC is used in [25] to design the IACG. The sliding mode control theory in [26] and [27] to solve the ITCG and IACG problems also contains a similar approaching law.

Inspired by the aforementioned research, this paper proposed a unified guidance design approach called free-time convergence error dynamics (FeTCED), whose convergence time is independent of the initial system condition and the control parameters. In addition, the FeTCED has fewer parameters and clearer physical meaning compared with other approaches. It is easy to meet the terminal constraints and performance requirements. Because the guidance law design is essentially a finite-time error-tracking problem mentioned before, FeTCED could be used to design various guidance laws with FeTC features. In this paper, the specific design process of the leading angle control, impact angle control, and impact time control guidance laws with FeTCED are given and the effectiveness is verified by numerical simulation.

The rest of this paper is organized as follows. Section 2 gives preliminaries about the missile guidance model. Section 3 presents the FeTCED and analyzes its convergence feature. Section 4 provides three illustrative guidance laws, including FeTCED-LACG, FeTCED-IACG, and FeTCED-ITCG. Finally, the simulation results and conclusions are offered in Sections 5 and 6.

## 2. Missile guidance model

Before presenting the guidance results, the following general assumptions, which are widely applied in guidance design, are given as follows: 1) The missile is an ideal particle model and the missile's autopilot is also an ideal process without control delay. 2) The earth's gravity is ignored here. 3) The guidance command is perpendicular to the missile's velocity vector, i.e. the missile speed is not changed by the guidance command. Here, the velocity is a vector while the speed is a scalar. Note that the above assumptions are widely used in the missile guidance law design.

A 2D planar guidance model is considered in this paper. As shown in Fig. 1, M and T represent the missile and the target in the inertial frame, which is denoted as $o$-$xy$. ($t_m$, $n_m$) and ($e_r$, $e_\theta$) are the velocity frame and the line-of-sight (LOS) frame, respectively. The time-varying velocity of the missile is denoted by $\mathbf{v}_m$, whose scalar form, $v_m$, denotes the missile speed. The notations $q$ and $\varphi_m$ denote the LOS angle and flight path angle in the inertial frame, respectively. The relative distance is denoted by $r$. The leading angle $\theta_m$ is defined as the angle from the velocity to the LOS.

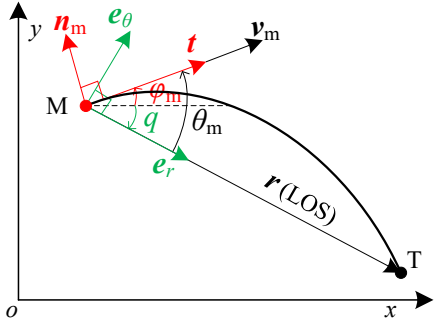

Fig. 1 Missile guidance model

The missile kinematic equations are given as

$$\begin{cases} \dot{x} = v_m \cos\varphi_m \\ \dot{y} = v_m \sin\varphi_m \end{cases} \quad (1)$$

where $x$ and $y$ are the postion of the missile in the inertial frame $o$-$xy$, and the dot symbol represents the derivation of variables to time. Besides, the relative kinematic equations between the missile and the target are listed below

$$\begin{cases} \dot{r} = -v_m \cos\theta_m \\ \dot{q} = -v_m \sin\theta_m / r \\ \dot{\varphi}_m = a_m / v_m \\ \varphi_m = q + \theta_m \end{cases} \quad (2)$$

where $a_m$ is the guidance acceleration to be designed.

The initial and terminal constraints of impact time and impact angle are as follows

$$\begin{cases} x(t_0) = x_0, y(t_0) = y_0, \varphi_m(t_0) = \varphi_0 \\ x(t_f) = x_f, y(t_f) = y_f, \varphi_m(t_f) = \varphi_f \\ t_f = t_d, \varphi_f = \varphi_d \end{cases} \quad (3)$$

where $t_0$ and $t_f$ represent the initial and terminal time respectively. $t_d, \varphi_d,$ and $\varphi_f$ are the desired impact time, desired impact angle, and termial impact angle respectively.

## 3. Free-Time convergence error dynamics

As presented in [18], the purpose of the guidance design is to solve finite-time error-tracking problems. The general form of error-tracking problems can be expressed as

$$\dot{\varepsilon}(t) = g(t)u(t) \quad (4)$$

where $\varepsilon(t)$ is the tracking error, $u(t)$ is the control command to be determined, and $g(t)$ represents a known function depending on the specific problems. For the missile guidance design problem, the tracking error could be selected as impact time error, impact angle error, zero-effort-miss, and so on. In general, $g(t) \neq 0$ for most guidance problems except for the ITCG. For the ITCG, $g(t)=0$ only at the equilibrium point, namely $\theta_m=0$. The singularity at the equilibrium point could be avoided by introducing an auxiliary function with higher-order infinitesimal at the equilibrium point[28].

For the guidance error tracking problems, the desired error dynamics, or reaching law in other words, is supposed to be determined first. A special non-autonomous differential equation was studied in [17]. The convergence feature of the differential equation is independent of the initial system conditions and the convergence time could be set freely. Combining the FeTC approach in [17] and the ED approach in [18], this paper gives free-time convergence error dynamics as **Theorem 1**.

**Theorem 1**. For the general error tracking problem in (4), supposing the desired error dynamics as

$$\begin{cases} \dot{\varepsilon}(t) + \dfrac{K}{T_s - t}\left(1 - e^{-\varepsilon(t)}\right) = 0, & \text{if } t \leq T_s \\ \dot{\varepsilon}(t) = 0, & \text{otherwise} \end{cases} \quad (5)$$

where $K \geq 1$, $0 < T_s < t_f$ is the convergence time that can be set freely and is independent of the initial system states and control parameters. Then the obtained control command can make the error converge to 0 at $T_s$.

**Proof.** Construct a Lyapunov function as

$$V(t) = \sqrt{\varepsilon^2} = |\varepsilon| \geq 0 \quad (6)$$

when $t \leq T_s$, differentiating $V(t)$ and substituting (5), we have

$$\dot{V} = \frac{\varepsilon \dot{\varepsilon}}{\sqrt{\varepsilon^2}} = -\frac{K\varepsilon}{V(T_s - t)}\left(1 - e^{-\varepsilon}\right)$$

$$\leq -\frac{K|\varepsilon|}{V(T_s - t)}\left(1 - e^{-|\varepsilon|}\right) \quad (7)$$

$$= -\frac{K}{T_s - t}\left(1 - e^{-V}\right)$$

The equation in (5) has the same form as the inequality in (7) when $t \leq T_s$. Let $x(t)$ be the solution of the following differential equation

$$\dot{x} + \frac{K}{T_s - t}\left(1 - e^{-x}\right) = 0 \quad (8)$$

Then we get

$$x(t) = \ln\left[C(T_s - t)^K + 1\right] \quad (9)$$

where the integral constant $C$ is determined by the initial parameter, namely $C = (e^{x_0} - 1)/(T_s - t_0)^K$, $x_0$ is the value of $x(t)$ at $t = t_0$.

Differentiating (9) further yields

$$\dot{x} = -\frac{CK(T_s - t)^{K-1}}{C(T_s - t)^K + 1} \quad (10)$$

From (9) and (10), when $t = T_s$, we have $\dot{x} = 0$ and $x = 0$. Therefore, for the Lyapunov function $V(t) \geq 0$, we have $\dot{V} \leq 0$ for all $t \leq T_s$, and the "=" holds if and only if $t = T_s$. In other words, when $t \rightarrow T_s$, the error $\varepsilon(t)$ and its derivation $\dot{\varepsilon}(t)$ tend to 0, and when $t \geq T_s$, $\varepsilon(t)$ remains 0. □

In the guidance law design, the magnitude of the initial leading angle error and the impact angle error is usually within $10^0$rad. The magnitude of the impact time error is usually within $10^1$s. As a result, the feature of FeTCED with an initial state $\varepsilon_0 < 10$ is analyzed below. The typical scenario of $\varepsilon_0 = 3$ is chosen in the following simulation.

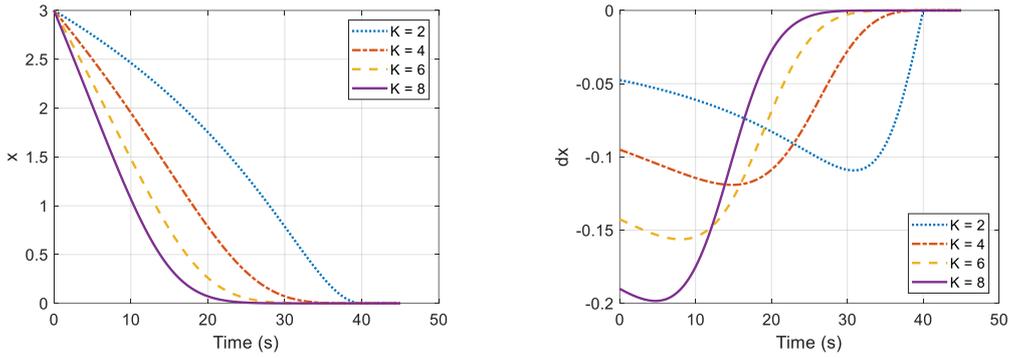

Fig. 2 Influence feature of K on (5)(Ts = 40s)

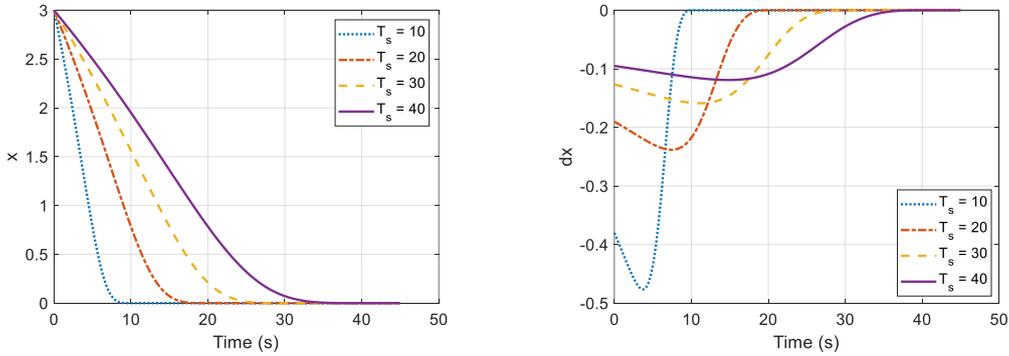

Fig. 3 Influence feature of $T_s$ on (5)($K = 4$)

Fig. 2 shows the influence feature of $K$ on the FeTCED. It can be seen from Fig. 2 that the larger the parameter $K$ is, the faster the state error $\varepsilon(t)$ converges, and the larger the corresponding initial value of $\dot{\varepsilon}(t)$ is when the free convergence time is set as $T_s$=40s. Besides, there is a trend of first increasing, then decreasing, and finally converging to 0 of $\dot{\varepsilon}(t)$. As a result, it is necessary to choose $K$ which makes the change of $\varepsilon(t)$ more gentle, and there will also be no sudden change of $u(t)$.

The influence feature of $T_s$ on the FeTCED is depicted in Fig. 3. When $K$=4, the larger the free convergence time $T_s$ is, the slower the convergence rate of $\varepsilon(t)$, and also the smaller the corresponding initial value of $\dot{\varepsilon}(t)$. Besides, the change curve of $\dot{\varepsilon}(t)$ is also smoother. Therefore, if $T_s$ is relatively small, the value of $K$ should be appropriately increased to make the change of $\dot{\varepsilon}(t)$ gentler, which is conducive to engineering realization.

In addition, the FeTCED approach has better performance when the initial error is small because the curve changes evenly. Therefore, it is necessary to convert the large error into the small error in the equivalent form. For example, in the general homing guidance law design process, the magnitude of the initial ZEM is $10^3$-$10^5$m generally, which could be converted into the equivalent leading angle or LOS rate. This feature is also reflected in the approximate value of Taylor expansion of $e^{\varepsilon(t)}$ in the next section later. On the other hand, the way to handle the tracking error in the guidance problem is to normalize the tracking error, which will be further studied in future works.

## 4. Illustrative guidance law design

In this section, the specific process of designing the corresponding guidance through the FeTCED approach is given by taking the ZEM, impact angle, and impact time into consideration, respectively.

### 4.1 FeTCED-LACG

The primary purpose of guidance law design is to minimize the ZEM for missiles against targets. When attacking a stationary target, ZEM has a one-to-one correspondence with the leading angle $\theta_m$, namely ZEM is equivalent to $\theta_m$. Consequently, the leading angle error is defined as

$$\varepsilon_\theta = \theta_m - \theta_d = \theta_m \quad (11)$$

Differentiating (11) and substituting (2) in, we have the leading angle error dynamics

$$\dot{\varepsilon}_\theta = \frac{a_m}{v_m} + \frac{v_m \sin \varepsilon_\theta}{r} \quad (12)$$

When $t \leq T_s$, selecting the desired FeTCED with respect to $\varepsilon_\theta$ as

$$\dot{\varepsilon}_\theta + \frac{K}{T_s - t}\left(1 - e^{-\varepsilon_\theta}\right) = 0 \quad (13)$$

where $K>1$ is a positive constant, and $T_s$ can be set freely.

Substituting (12) into (13) yields

$$a_m = -Kv_m \frac{1 - e^{-\varepsilon_\theta}}{T_s - t} - \frac{v_m^2 \sin \varepsilon_\theta}{r} \quad (14)$$

When $t>T_s$, the leading angle error converges to 0 already, namely $a_m$=0. Therefore, the FeTCED-LACG becomes

$$a_m = \begin{cases} -Kv_m \dfrac{1 - e^{-\varepsilon_\theta}}{T_s - t} - \dfrac{v_m^2 \sin \varepsilon_\theta}{r} &, t \leq T_s \\ 0 &, t > T_s \end{cases} \quad (15)$$

For the FeTCED-LACG, if the convergence time $T_s$ is set to the impact time $t_f$, there is no case of $t>T_s$. According to the linearization in [8], when $\theta_m$ is small, we have

$$T_s = t_f \approx \frac{r}{v_m} + t \quad (16)$$

and

$$\sin \varepsilon_\theta \approx \varepsilon_\theta = \theta \quad (17)$$

Then the Taylor expansion is carried out for $e^{-\varepsilon_\theta}$ near $\varepsilon_\theta=0$, and the first two terms are taken as approximations, we get

$$e^{-\varepsilon_\theta} \approx 1-\varepsilon_\theta = 1-\theta \quad (18)$$

Besides, the LOS rate can be approximately expressed as

$$\dot{q} \approx -\frac{\theta_m}{t_{go}} = -\frac{\theta_m}{T_s-t} = -\frac{\theta_m}{r/v_m} \quad (19)$$

Substituting (2), (18), and (19) into (15) yields

$$a_m = -Kv_m \frac{1-e^{-\varepsilon_\theta}}{T_s-t} - \frac{v_m^2 \sin\varepsilon_\theta}{r} \quad (20)$$
$$\approx (K+1)v_m\dot{q} = Nv_m\dot{q}$$

where $N=K+1$. This is the classic PNG. The performance of PNG is analyzed in detail in [29]-[35]. PNG has the advantages of simple form, strong robustness, and easy implementation. Motivated by the BPNG, the IACG and ITCG are designed based on the FeTCED approach by taking the impact angle constraint and impact time constraint into consideration respectively.

## 4.2 FeTCED-IACG

In order to increase survivability and destructiveness, the impact angle is often considered in the guidance law design. The proposed FeTCED approach is applied to develop a novel IACG, which is a kind of improved BPNG. The impact angle can be controlled to the desired value by adding a biased term based on the PNG. The form of IACG is given as

$$a_m = a_{PNG} + a_{IA} \quad (21)$$

where $a_{PNG}$ is the PNG term and $a_{IA}$ is a biased term to regulate the terminal impact angle.

According to [33], when the missile is guided by PNG, the terminal impact angle can be expressed as

$$\varphi_f = \frac{N}{N-1}q - \frac{1}{N-1}\varphi_m \quad (22)$$

Note that $\theta_m$ will converge to 0 eventually, namely $\varphi_f=q_f$.

Let the desired terminal impact angle be $\varphi_d$, then the impact angle error could be defined as

$$\varepsilon_\varphi = \varphi_d - \varphi_f \quad (23)$$

To achieve zero impact angle error, taking the derivative of $\varepsilon_\varphi$ and substituting (2), (21), and (22) in, sorting and simplifying, we have the impact angle error dynamics as follows

$$\dot{\varepsilon}_\varphi = \frac{1}{(N-1)v_m}a_{IA} \quad (24)$$

When $t \leq T_s$, selecting the desired FeTCED with respect to $\varepsilon_\varphi$ as

$$\dot{\varepsilon}_\varphi + \frac{K}{T_s-t}\left(1-e^{-\varepsilon_\varphi}\right) = 0 \quad (25)$$

Substituting (24) into (25) yields

$$a_{IA} = \frac{K(N-1)v_m}{T_s-t}\left(1-e^{-\varepsilon_\varphi}\right) \quad (26)$$

When $t>T_s$, the impact angle error converges to 0 already, namely $a_{IA}=0$. Combining (21) and (26), we have the FeTCED-IACG as follows

$$a_m = \begin{cases} Nv_m\dot{q} + \frac{K(N-1)v_m}{T_s-t}\left(1-e^{-\varepsilon_\varphi}\right), & t \leq T_s \\ Nv_m\dot{q}, & t > T_s \end{cases} \quad (27)$$

From **Theorem 1**, under the FeTCED-IACG, the impact angle error $\varepsilon_\varphi$ converges to 0 at $T_s$. After that, the FeTCED-IACG reduces to the PNG law.

Similar to the previous subsection, Taylor expansion is carried out for $a_{IA}$ near $\varepsilon_\varphi=0$, and the first two terms are taken as approximations. Besides, $T_s$ is set at the terminal impact time $t_f$. Then (27) is converted into

$$a_m = Nv_m\dot{q} - \frac{K(N-1)v_m}{t_{go}}\varepsilon_\varphi \quad (28)$$

The guidance law shown in (28) is the OED-based impact angle control guidance (OED-IACG) proposed in [18]. The result also shows that when the impact angle error is small, the closer $T_s$ is to $t_f$, the more similar the performance of FeTCED-IACG is to that of OED-IACG.

### 4.3 FeTCED-ITCG

In order to increase the lethality and penetration probability, a salvo attack with multiple missiles becomes a hot research field to attack a protected target, such as a ship with close-in weapon systems and important infrastructure with anti-missile weapon systems. One typical guidance law to achieve a salvo attack is to design the ITCG law with the same desired impact time for each missile. Similar to the previous subsection, the ITCG consists of two parts: the PNG term to achieve zero ZEM and the biased term to nullify the impact time error, namely

$$a_m = a_{PNG} + a_{IT} \qquad (29)$$

where $a_{IT}$ is the biased term to control impact error.

From [2], the total impact time estimation under the PNG law is given as

$$t_f \approx t + \frac{r}{v_m}\left(1 + \frac{\theta_m^2}{2(2N-1)}\right) \qquad (30)$$

where $t$ is the current flying time. Denoting the desired impact time as $t_d$. Then the impact time error can be defined as

$$\varepsilon_t = t_d - t_f \qquad (31)$$

Differentiating (31), and substituting (2), (29), and (30) in, we have

$$\begin{aligned}\dot{\varepsilon}_t &= -\frac{\dot{r}}{v_m}\left(1 + \frac{\theta_m^2}{2(2N-1)}\right) - \frac{r\theta_m\dot{\theta}_m}{(2N-1)v_m} - 1 \\ &= \cos\theta_m\left(1 + \frac{\theta_m^2}{2(2N-1)}\right) + \frac{(N-1)\theta_m\sin\theta_m}{2N-1} \\ &\quad - \frac{r\theta_m}{(2N-1)v_m^2}a_{IT} - 1\end{aligned} \qquad (32)$$

Taking the assumption of small leading angle into consideration, namely, $\theta_m$ is small, then we have the approximation as

$$\sin\theta_m \approx \theta_m, \quad \cos\theta_m \approx 1 - \frac{\theta_m^2}{2} \qquad (33)$$

Substituting (33) into (32), neglecting the higher-order terms, sorting and simplifying, and then the impact time error dynamics can be obtained as

$$\dot{\varepsilon}_t = -\frac{r\theta_m}{(2N-1)v_m^2}a_{IT} \qquad (34)$$

When $t \leq T_s$, selecting the desired FeTCED with respect to $\varepsilon_t$ as

$$\dot{\varepsilon}_t + \frac{K}{T_s - t}\left(1 - e^{-\varepsilon_t}\right) = 0 \qquad (35)$$

Substituting (34) into (35) yields

$$a_{IT} = \frac{K(2N-1)v_m^2}{r\theta_m(T_s - t)} \cdot \left(1 - e^{-\varepsilon_t}\right) \qquad (36)$$

When $t > T_s$, the impact time error converges to 0 theoretically, namely $a_{IT}=0$. Combining (29) and (36), we have the FeTCED-ITCG as follows

$$a_m = \begin{cases} Nv_m\dot{q} + \dfrac{K(2N-1)v_m^2}{r\theta_m(T_s-t)}\left(1-e^{-\varepsilon_t}\right), & t \leq T_s \\ Nv_m\dot{q}, & t > T_s \end{cases} \qquad (37)$$

From **Theorem 1**, under the FeTCED-ITCG, the impact time error $\varepsilon_\varphi$ converges to 0 at $T_s$. After that, the FeTCED-ITCG reduces to the PNG law.

Similar to the previous subsection, taking Taylor expansion for $a_{IT}$ near $\varepsilon_t=0$, and the first two terms are taken as approximations. Besides, $T_s$ is set at the terminal impact time $t_f$. Then the FeTCED-ITCG shown in (37) is converted into the OED-based impact time control guidance (OED-ITCG) proposed in [18] as follows

$$a_m = Nv_m\dot{q} + \frac{K(2N-1)v_m^2}{r\theta_m t_{go}}\varepsilon_t \qquad (38)$$

Note that the magnitude of impact time error is usually within $10^1$s, not always a small value that is close to 0. If the chosen impact time error is small in the specific guidance scenario, the closer $T_s$ is to $t_f$, the more similar the performance of FeTCED-ITCG is to that of OED-ITCG.

## 5. Simulation results

The guidances designed in the previous section are analyzed and demonstrated by numerical simulation in this section. Numerical simulations are conducted in an air-to-ground engagement scenario. In all the following simulations, the target is located at the origin of the reference frame. The initial relative distance between the missile and the target is 20km with an initial LOS angle of -45°. The flying speed of the missile is 500m/s with an initial flying path angle of 0°. That is to say, the missile is flying horizontally. The simulation step is 0.01s. In order to better analyze the guidance performance of the designed guidance laws, the command saturation and the blind area condition of the seeker are not set. the performance of the three designed guidance laws is analyzed below in this scenario.

### 5.1 Analysis of FeTCED-LACG

As shown in (20), when $T_s$ is set at the terminal impact time $t_f$, the FeTCED-LACG can be approximated as the PNG with $N=K+1$. As a result, the FeTCED-LACG with different free convergence times is compared with PNG. The guidance parameters of FeTCED-LACG and PNG are chosen in Table 1. The guidance laws of Missile 1~3 are the FeTCED-LACG, and the guidance parameter is chosen as $K=3$, the convergent time is set as 20s, 30s, and 40s respectively. The guidance law of Missile 4 is PNG with $N=4$. The simulation results of the missile trajectory, acceleration command, leading angle error, and energy cost are given in Fig. 4.

Table 1 Parameters of FeTCED-LACG and PPN

| No. | Guidance Law | $K$ | $T_s$ (s) |
| --- | --- | --- | --- |
| Missile 1 | FeTCED-LACG | 3 | 20 |
| Missile 2 | FeTCED-LACG | 3 | 30 |
| Missile 3 | FeTCED-LACG | 3 | 40 |
| Missile 4 | PNG | 4 | \ |

Fig. 4 (a) shows the trajectories of 4 missiles that the larger $T_s$, the more curved the missile trajectory. The acceleration command and leading angle error are presented in Fig. 4 (b) and Fig. 4 (c). It can be seen that the acceleration command and leading angle error of Missile 1-3 converge to 0 within the set time. Besides, the larger $T_s$, the smaller the initial acceleration command, and the slower the change of leading angle error. Although the acceleration command curve of Missile 4 guided by PNG is smoother than that of Missile 1-3, the leading angle error converges to 0 at the terminal impact time. As can be seen from Fig. 4 (d), the energy cost decreases with the increase of $T_s$. When $T_s$ is set close to the terminal impact time $t_f$, the energy cost of the FeTCED-LACG is less and the performance is closer to PNG.

### 5.2 Analysis of FeTCED-IACG

Given space limitations, this paper will not analyze the performance of FeTCED-IACG under different desired impact angles or guidance parameters, but only choose the typical case where the desired impact angle is $\varphi_d=-90°$. Besides, it can be seen from (27) and (28) that when the convergent time is set as the impact time, namely $T_s=t_f$, the FeTCED-IACG can be approximated as the OED-IACG. Therefore, the FeTCED-IACG and OED-IACG under different free convergent time settings are compared and analyzed in this subsection. The guidance parameters of FeTCED-IACG and OED-IACG are chosen in Table 2. The guidance laws of Missile 1~3 are the FeTCED-IACG and that of Missile 4 is the OED-IACG. The guidance parameters are all chosen as $N=4$ and $K=3$. The convergent time of Missile 1~3 is set as 20s, 30s, and 40s respectively. The simulation results are provided in Fig. 5.

Table 2 Parameters of FeTCED-IACG and OED-IACG

| No. | Guidance Law | $(N,K)$ | $T_s$(s) |
| --- | --- | --- | --- |
| Missile 1 | FeTCED-IACG | (4, 3) | 20 |
| Missile 2 | FeTCED-IACG | (4, 3) | 30 |
| Missile 3 | FeTCED-IACG | (4, 3) | 40 |
| Missile 4 | OED-IACG | (4, 3) | \ |

It can be seen from Fig. 5 (a) that all 4 missiles can destroy the target successfully at the desired impact angle. However, from Fig. 5 (b), the initial acceleration and the maximum acceleration of Missile 4 guided by OED-IACG are smaller, compared with that of the other 3 missiles guided by FeTCED-IACG. For FeTCED-IACG, the larger the free convergence time is set, the smaller the initial acceleration is, and also the closer the missile trajectory is to that of the OED-IACG, which also verifies the conclusion that the closer the free convergence time is to the total impact time, the closer the performance of the FeTCED-IACG is to the OED-IACG when the impact angle error is small. The impact angle error is depicted in Fig. 5 (c). The impact angle error of Missile 1~3 guided by FeTCED-IACG can converge to 0 within the desired free convergence time, while that of Missile 4 converges to 0 only at the impact time. Fig. 5 (d) presents the energy cost curve of 4 missiles. The energy cost of Missile 4 is the least, and the energy cost of Missile 1-3 decreases with the increase of the free convergence time.

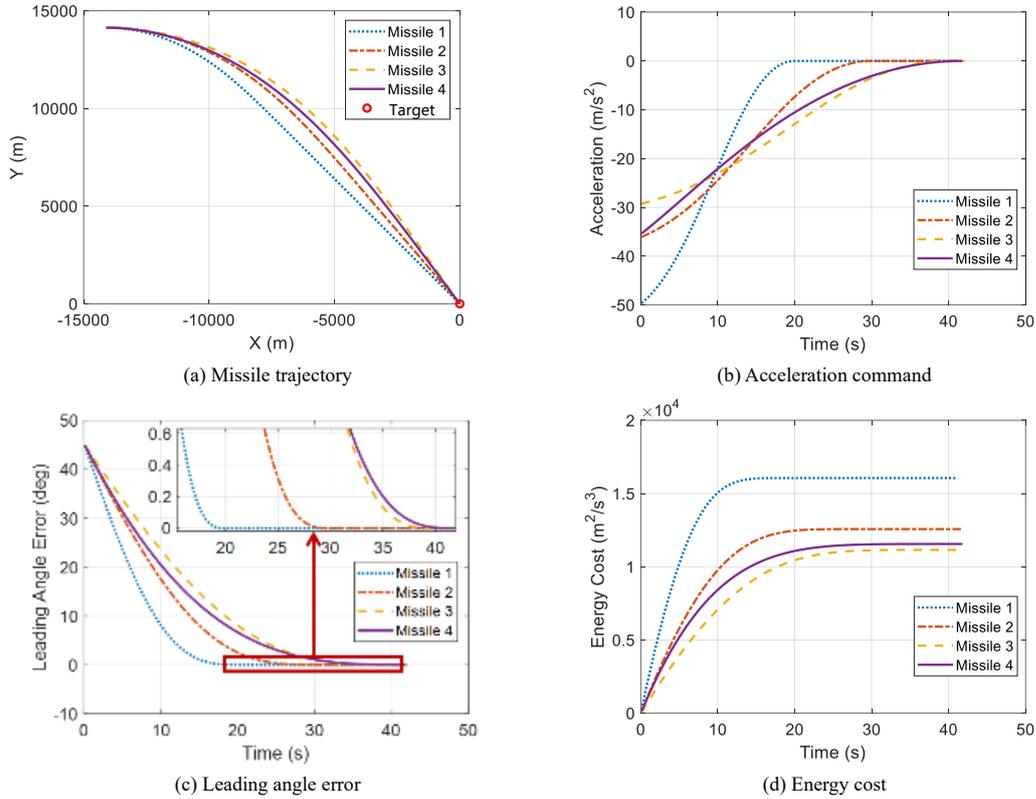

(a) Missile trajectory  (b) Acceleration command

(c) Leading angle error  (d) Energy cost

Fig. 4 Results of FeTCED-LACG and PNG

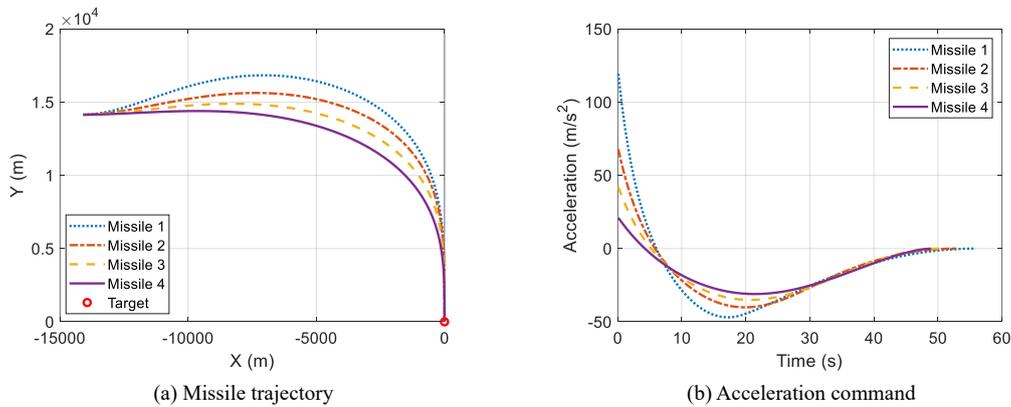

(a) Missile trajectory  (b) Acceleration command

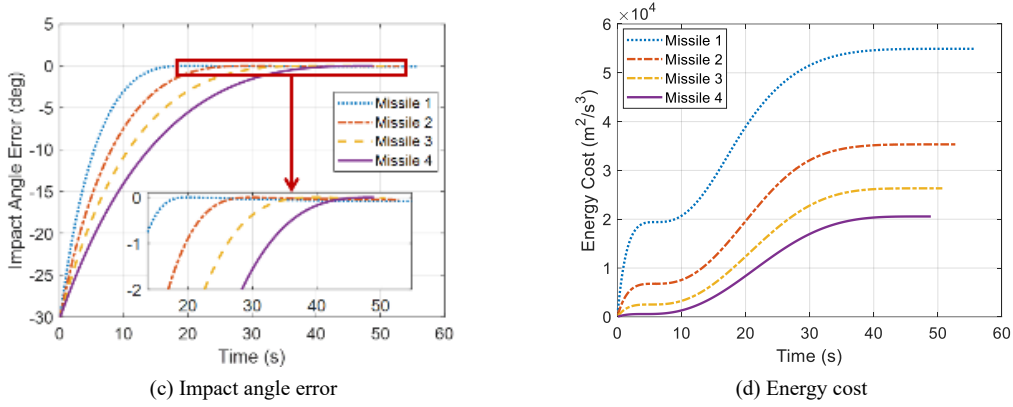

(c) Impact angle error  (d) Energy cost

Fig. 5 Results of FeTCED-IACG and OED-IACG

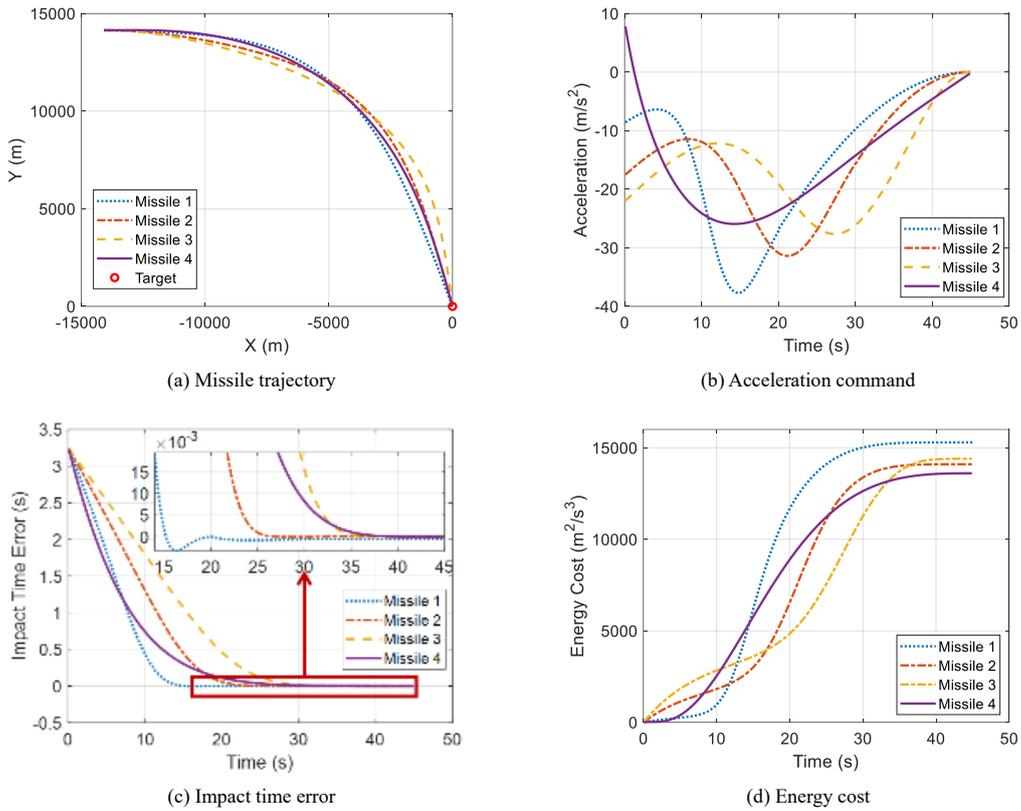

(a) Missile trajectory  (b) Acceleration command

(c) Impact time error  (d) Energy cost

Fig. 6 Results of FeTCED-ITCG and OED-ITCG

### 5.3 Analysis of FeTCED-ITCG

According to the initial relative information of the missile and the target, the total impact time of the missile guided by PNG is about 41.8s estimated by (30). From (37) and (38), when the convergence time is set as the terminal impact time and the error is small, the performance of FeTCED-ITCG is similar to that of OED-ITCG. Similar to the previous subsection, the typical scenario where the desired impact time is 45s is chosen to compare FeTCED-ITCG and OED-ITCG. The guidance parameters of FeTCED-ITCG and OED-ITCG are chosen in Table 3 and the simulation results are presented in Fig. 6.

From Fig. 6 (a) and Fig. 6 (c), it can be seen that all 4 missiles attack the target successfully in the desired impact time. The acceleration command of Missile 4 guided by OED-ITCG is small in the initial phase and the energy cost is the least, as provided in Fig. 6 (b) and Fig. 6 (d).

In the middle phase, the acceleration command first increases and then decreases. The energy consumption also increases rapidly, but the total energy cost is still the least, which is consistent with the optimality of OED-ITCG. However, its impact time error converges to 0 at the terminal impact time. For FeTCED-ITCG, it can be seen from Fig. 6 (c) that the impact time errors of the three missiles converge to 0 at the given time, but the corresponding energy cost is also more than that of OED-ITCG.

Table 3 Parameter of FeTCED-ITCG and OED-ITCG

| No. | Guidance Law | (N, K) | Ts(s) |
| --- | --- | --- | --- |
| Missile 1 | FeTCED-ITCG | (4, 5) | 20 |
| Missile 2 | FeTCED-ITCG | (4, 5) | 30 |
| Missile 3 | FeTCED-ITCG | (4, 5) | 40 |
| Missile 4 | OED-ITCG | (4, 5) | \ |

# 6.Conclusions

This paper proposed a unified free-time convergence guidance law design approach. The difference between the finite-time convergence approach and the fixed-time convergence approach is compared and analyzed. Based on the free-time convergence theory and error dynamics approach, the FeTCED approach is proposed to solve error-tracking problems whose convergent time can be set independent of the initial system state and control parameters. Using this approach, the guidance laws of leading angle control, impact angle control, and impact time control with free-time convergence feature are derived as examples respectively. The cons and pros of the proposed guidance laws and the existing similar guidance laws are compared and analyzed through numerical simulation.

The guidance design approach, which is the combination of the prediction-correction method and free-time convergence method, has great engineering application potential for its simple structure, clear physical meaning, and few adjustable parameters, which is the main contribution of the paper. Besides, the method proposed can be adapted to handle more guidance scenarios, such as guiding multiple missiles to intercept a maneuvering target or guiding a missile while considering multiple constraints.